\documentclass[12pt]{article} 

\usepackage[numbers,sort&compress]{natbib}
\newcommand{\stitle}{Precipitates in high-pressure torsion treated Mg-3.4at.\%Zn alloy}

\usepackage[left=2.5cm,right=2.5cm,top=2cm,bottom=2cm]{geometry}
\usepackage{fancyhdr}
	\fancypagestyle{plain}{ %
	\fancyhf{} 
	\lhead{\textit{\stitle}}
	\rhead{Pauw \textit{et al.}}
	\lfoot{\textit{Preprint}}
	\rfoot{\thepage}
	}

\usepackage{amsbsy} 
\usepackage{graphicx}
	\graphicspath{{Figures/}} 
\usepackage{subfigure}
\usepackage[version=3]{mhchem} 
\usepackage{multirow} 
\usepackage{multicol}
\usepackage{booktabs}	
\usepackage{setspace}
\usepackage{xspace}
\usepackage{comment}
\usepackage{rotating}
\usepackage{layout}

\usepackage[plainpages=false,
pdftitle={\stitle},
pdfauthor={Pauw, Rosalie, Meng, Somekawa, Mamiya, Kitazawa, Tsuchiya},
pdfsubject={\stitle},
pdfkeywords={Keywords},
colorlinks,
debug,
linkcolor=red,
citecolor=red, 
filecolor=red,
breaklinks=false]{hyperref}
\usepackage{doi} 

\title{Characterisation of precipitates formed in high-pressure torsion treated Mg-3.4at.\%Zn alloy}

\author{B. R. Pauw$^{a}$,
\and J. M. Rosalie$^{a*}$
\and F. Meng$^{a,b}$
\and H. Mamiya$^{a}$
\and H. Kitazawa$^{a}$
\and K. Tsuchiya$^{a,c}$
}	
\date{}
\vspace{6pt}

\begin{document}

\maketitle
\noindent
$^a$National Institute for Materials Science (NIMS), Japan.\\
$^b$Ames Laboratory, U.S. Department of Energy, Ames, Iowa 50011, USA.\\
$^c$University of Tsukuba. Japan.\\
\begin{abstract} 
Microstructural analysis of a Mg-Zn alloy deformed at room temperature by high-pressure torsion (HPT) indicates that fine-scale precipitation occurs even without post-deformation heat treatment. Small-angle X-ray scattering detects precipitates with radii between 2.5--20\,nm after one rotation, with little increase in particle size or volume fraction after 20  rotations. High resolution electron micrographs identify grain boundary precipitates of  monoclinic \ce{Mg4Zn7} phase after three  rotations and \ce{MgZn2} after 20 rotations.
\end{abstract}

\medskip
\section{Introduction}

Severe plastic deformation (SPD) comprises a suite of techniques which impose high plastic strain on a material in order to reduce the grain size to the micron or nanometer regime. Such grain refinement substantially increases the yield stress via the Hall-Petch effect, while simultaneously improving the ductility \cite{Yamashita2001,DingChang2009,LiWei2011}, and can lead to high-strain rate superplasticity \cite{Komura1998}. This desirable combination of properties has generated great interest in SPD techniques, and particularly in equal-channel angular extrusion (ECAE) and high-pressure torsion (HPT) which impose high strains with no net shape change. 

The ultrafine grain sizes that develop during HPT are inherently thermodynamically unstable \cite{WangIwahashi1996} and restriction of grain growth during and after processing is important in order to retain the improved mechanical properties. Grain growth occurs more rapidly and at lower temperatures than usual due to the high defect densities and the resulting ultra-fast atomic diffusion that occurs during SPD \cite{Amouyal2007,Divinski2010,Ribbe2009a}. During HPT, dynamic grain growth occurs even without external heating and results in ``saturation'' where the rate of dynamic grain growth balances that of strain-induced recrystallisation and a further increase in strain will not lead to a reduction of grain size.

One approach used to restrict grain growth is the use of fine precipitates to pin grain boundaries. For example, fine-scale precipitation of Al-Sc particles has been shown to reduce the rate of grain growth of ECAE Al-Mg-Sc \cite{Komura1998} leading to high-strain rate superplasticity. More recently it was demonstrated that grain sizes of approximately 140\,nm could be achieved in HPT-deformed Mg-Zn \cite{MengRosalie2014}. This is a much more refined grain size than in pure Mg, which reaches saturation at a grain size of  $\sim1\,\mu$m \cite{EdalatiHpt2011}. The difference in grain size was attributed to the presence of fine-scale zinc-rich grain boundary precipitates that were observed in high angle annular dark field (HAADF) STEM images after 20 revolutions of HPT deformation.

In this work we identify the precipitates formed during HPT of a Mg-Zn alloy using high resolution transmission electron microscopy (HRTEM), and quantify the precipitate size distribution using small angle X-ray scattering (SAXS). The complementary TEM and SAXS methodology has recently been shown to be a valuable tool for quantifying particle size and volume fraction in similar alloys \cite{RosaliePauw2014} and here we demonstrate its applicability to studies of  precipitation in SPD materials.

\section{Experimental details}

The experiments were conducted on disc-shaped samples (10\,mm diameter, $\sim$0.85\,mm height) cut in cross-section from an extruded rod of a Mg-3.4at.\%Zn alloy (The alloy casting and prior heat treatment is described in detail by Rosalie and Pauw \cite{RosaliePauw2014}). Discs were solution-treated at 300$^\circ$C for 1\,h in an Ar atmosphere and quenched in ambient temperature water. They were subsequently deformed at room temperature by HPT. Each disc was compressed with an applied pressure of 5\,GPa. This load was maintained for 30\,s before and during the application of rotational strain at a rotation speed of 1\,RPM. Previous studies on a \ce{Zr50(Cu , Al)50} alloy under similar deformation conditions found that frictional heating only  increased the anvil temperature adjacent to the sample to $\sim$306\,K \cite{MengTsuchiya2013}. The temperature increase for the present material is expected to be significantly less and insufficient to facilitate precipitation during HPT. The height of the discs after HPT treatment was $\sim$0.55\,mm.

TEM foils were prepared from two HPT discs, one of which was subjected to 3\,HPT rotations (N=3), and the second subjected to 20 rotations (N=20). These conditions were selected as N$>$1 was required to give a homogeneous hardness distribution and microstructure \cite{MengRosalie2014}, thus ensuring that the TEM foils are representative of the disc as a whole. The N=3 condition corresponds to the peak hardness condition, whereas the N=20 sample was in the strain-saturated condition \cite{MengRosalie2014}. The samples were prepared for TEM by mechanical grinding and polishing, followed by dimple grinding and thinning to perforation by precision ion polishing. Foils were examined using JEOL 2100 and 2100F microscopes, operating at 200\,kV. The precipitate phases were identified from Fast Fourier transforms of high-resolution images. A Hann filter was used to remove high frequency noise from the region of interest before performing the FFT.

SAXS analysis was performed on samples in four different states: one sample preceding the HPT treatment (i.e. solution-treated, ``ST''), one where the sample was compressed but not rotated (``Compressed'', or N=0), one sample after 1\,HPT rotation (N=1), and one at N=20. The samples were thinned to approximately 0.1\,mm, and measured for 21600\,s each. SAXS measurements were performed on a Bruker Nanostar instrument with a chromium target producing 5.4\,keV photons, and a sample-to-detector distance of 1.05m, resulting in a Q-range coverage of $0.05\leq$Q$\leq 1.3\, \mbox{nm}^{-1}$ (with $Q=4\pi/\lambda \sin(\theta)$, where $2\theta$ denotes the scattering angle, and $\lambda$ the X-ray photon wavelength). A photon-counting wire array (delay line) detector was used to collect the scattered photons.

Collected images were corrected for image distortion using Bruker's ``2D SAXS'' software supplied with the instrument. The following corrections were subsequently performed using an in-house developed data correction procedure (see \cite{Pauw2013a} for the details of each correction step): data read-in corrections (DS), darkcurrent (DC), pixel masking (MK), flatfield (FF), time (TI), flux (FL), transmission (TR), sample self-absorption (SA), spherical distortion (SP), background (BG), and thickness (TH), followed by a data integration in logarithmically-spaced Q-bins. Uncertainty estimates, used as weights in the fitting procedure, have been determined as the largest of either the propagated Poisson uncertainties, 1\% of the intensity in the bin, or the standard error in the bin. 

After correction, the data was subjected to a Monte Carlo-based size distribution determination procedure assuming spherical scatterers \cite{Pauw2013}, whose size (radii) range is defined by the Q-range to between 2.5 and 60\,nm. On average, the scattering patterns fit to within the uncertainty of the data over the entire Q-range (i.e. to $\chi^2_r \leq 1$). 

\section{Results and discussion}

\subsection{TEM}

\begin{figure}
\begin{center}
	\begin{center}
	\makebox[2ex]{}
	\hfill
	TEM
	\hfill\
	\hfill
	HRTEM
	\hfill\
	\hfill
	FFT
	\hfill\
	\hfill\
	\end{center}
	
	\raisebox{0.15\textwidth}[0pt][0pt]{
		\begin{rotate}{90}
			N=3
		\end{rotate}}
	\hfill
       \subfigure[\label{hrtem-15}]{\includegraphics[width=0.3\textwidth]{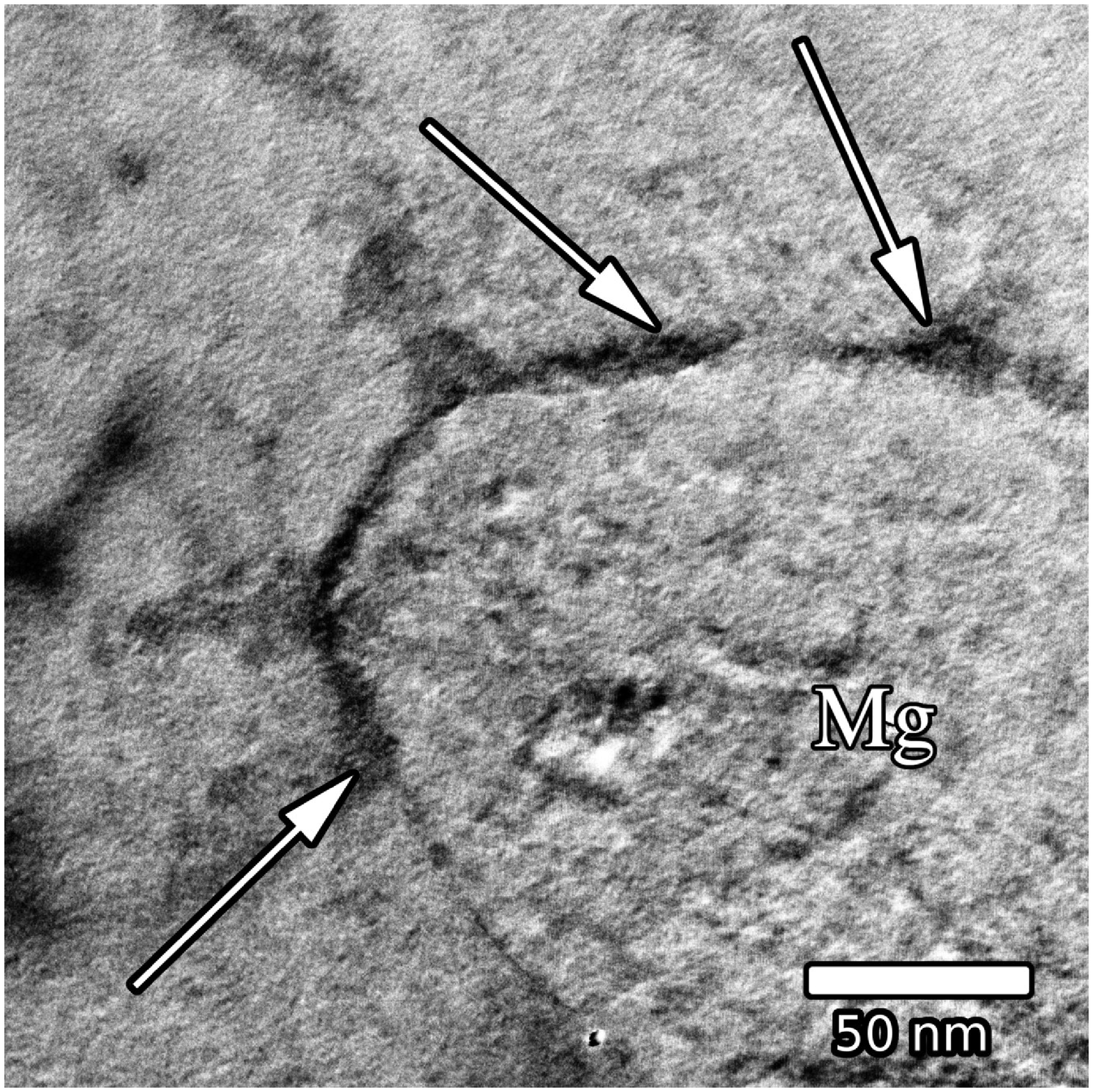}}\hfill
       \subfigure[ \label{hrtem-23-a}]{\includegraphics[width=0.3\textwidth]{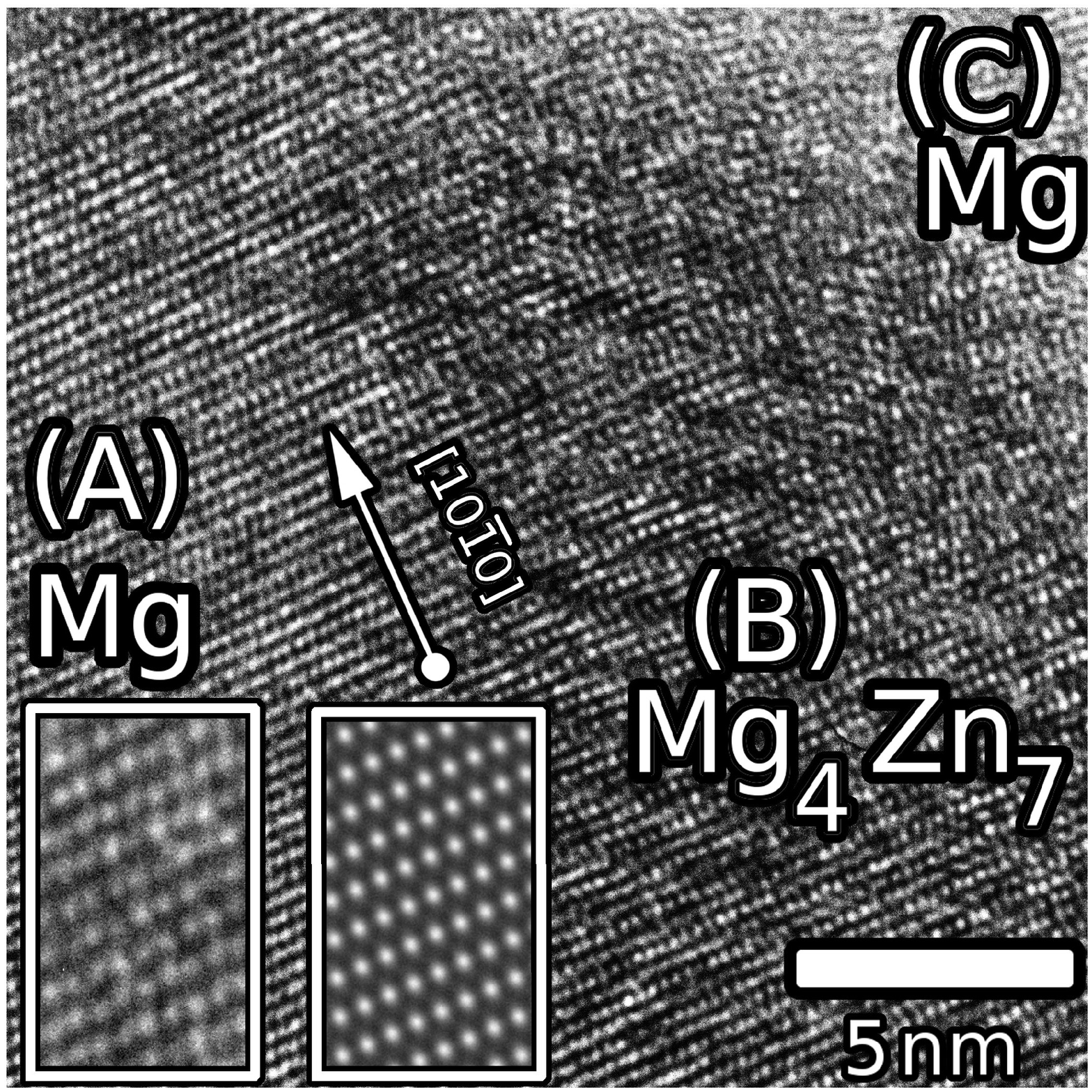}}
	\hfill
	\subfigure[\label{hrtem-23-fft-Mg4Zn7}]{\includegraphics[width=0.3\textwidth]{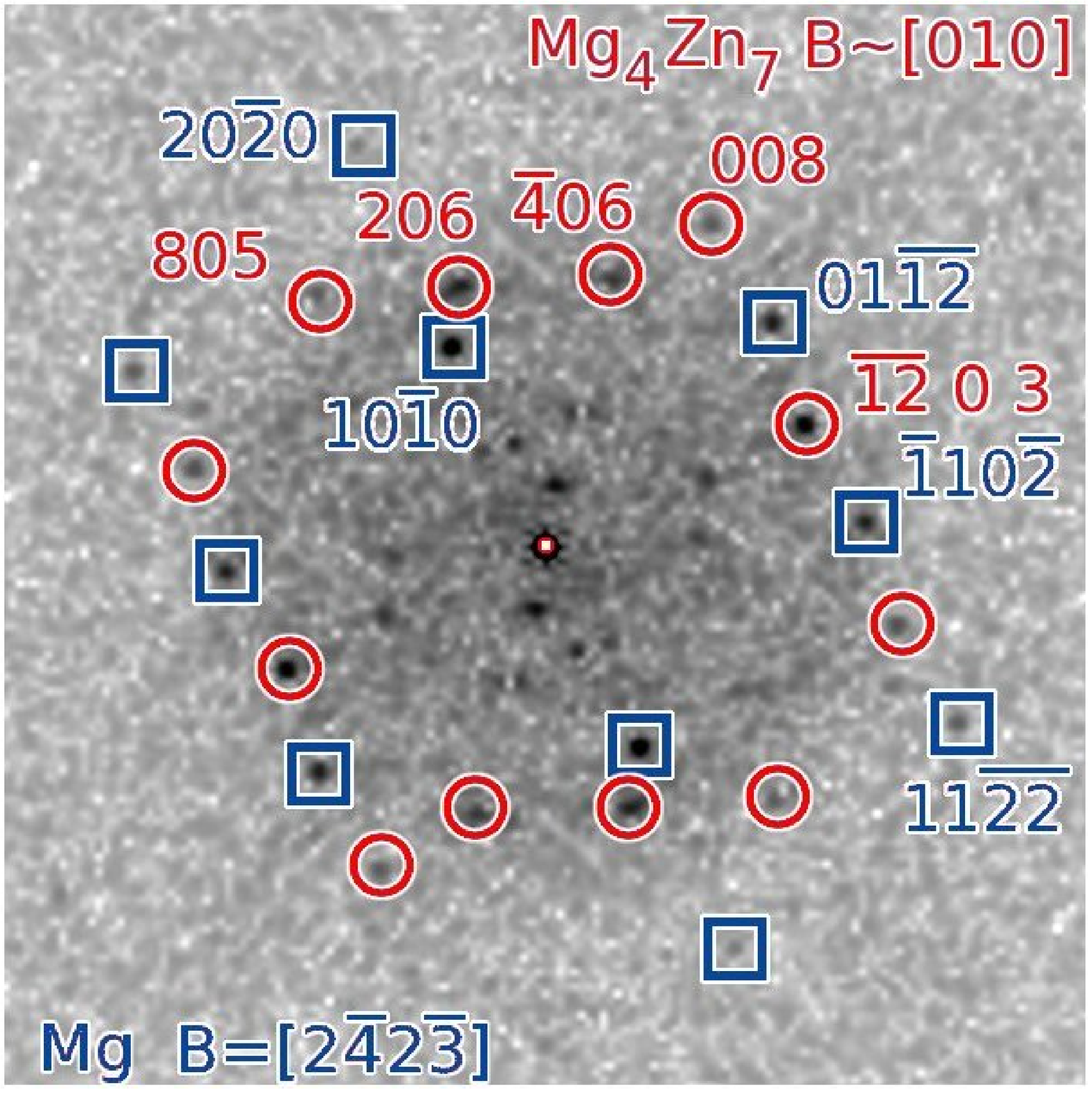}}
	\hfill\

	\raisebox{0.15\textwidth}[0pt][0pt]{
		\begin{rotate}{90}
			N=20
		\end{rotate}}
	\hfill
\subfigure[\label{N20_10}]{\includegraphics[width=0.3\textwidth]{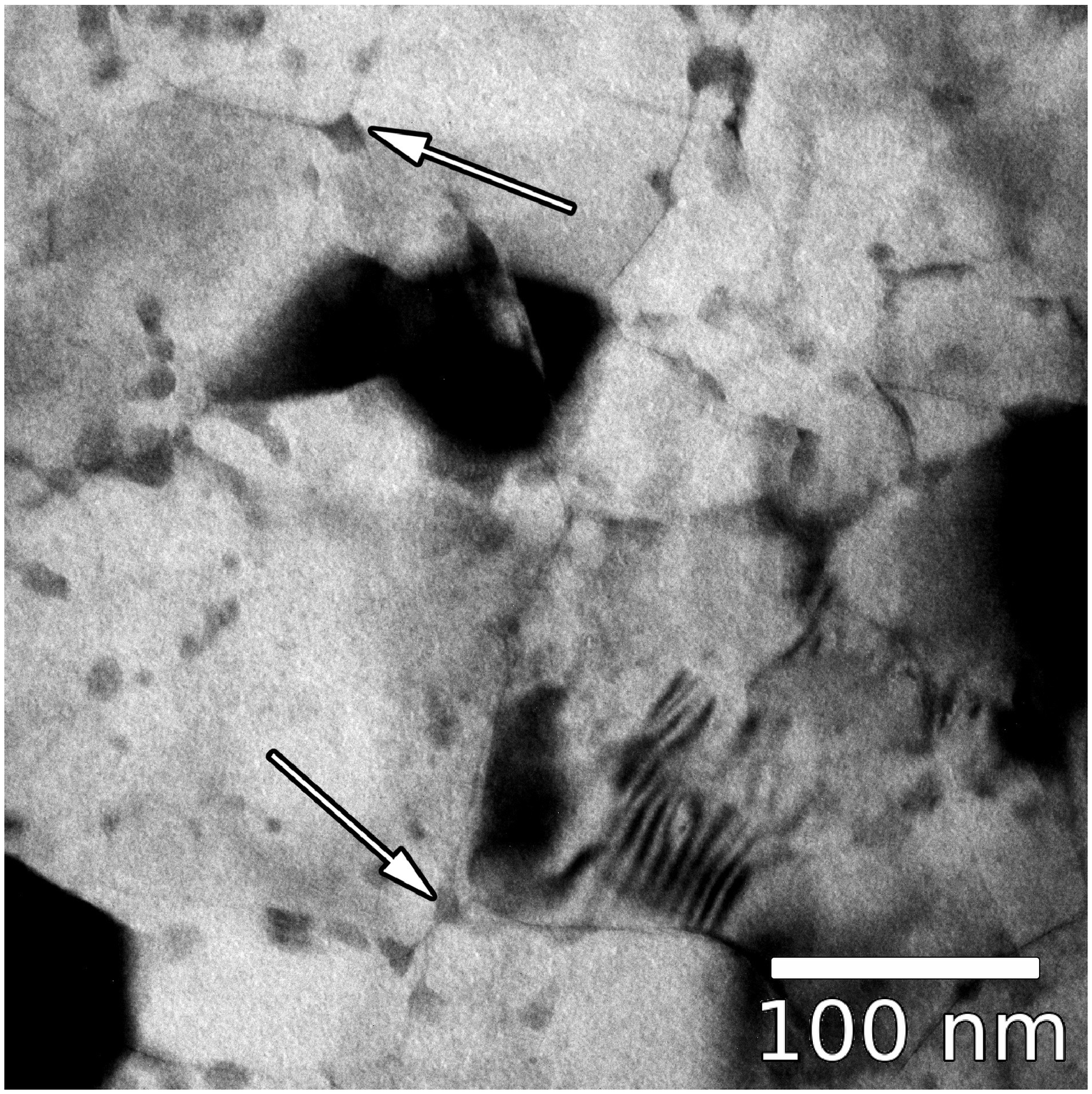}}
	\hfill
	\subfigure[\label{hrtem-n20}]{\includegraphics[width=5cm]{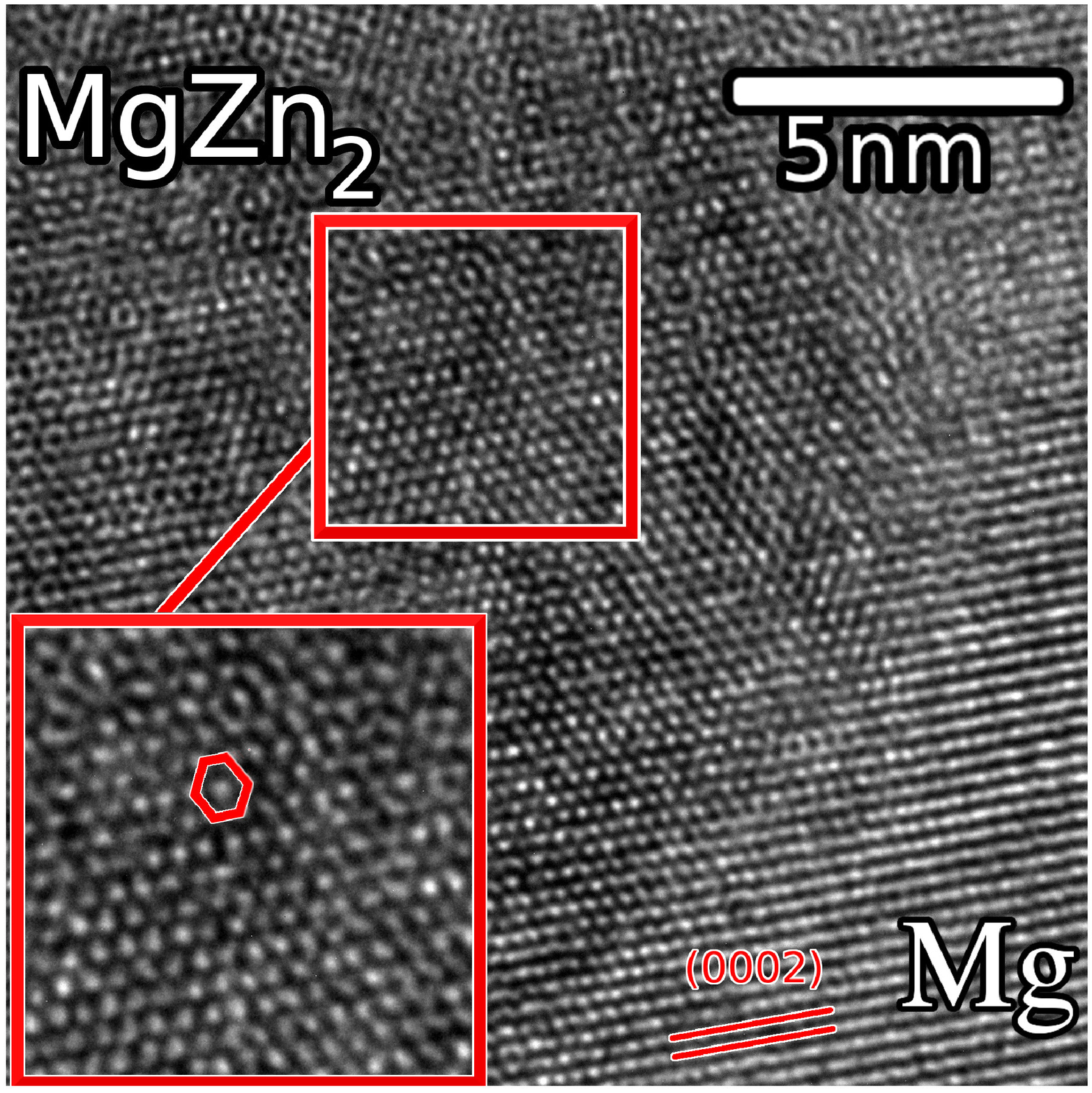}}
	\hfill
	\subfigure[\label{hrtem-n20-fft}]{\includegraphics[width=5cm]{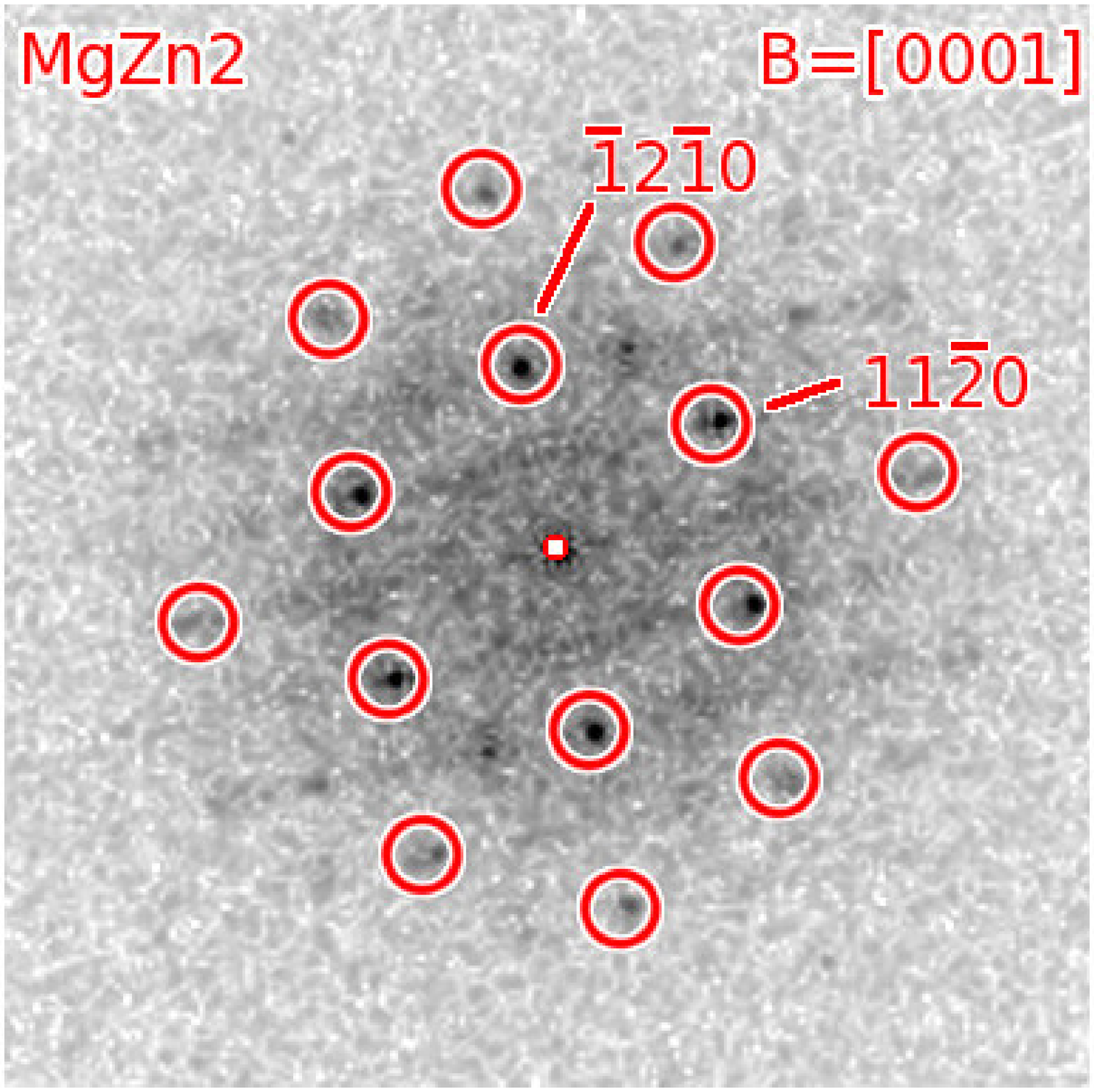}}
	\hfill\

	\caption{Electron micrographs of grain boundary precipitates in HPT Mg-Zn. Figures (a)--(c) are from samples deformed to N=3 and (d)--(f) are for N=20. (a,d) show typical diffraction contrast images of grain boundary precipitation. High resolution images are presented in Figs.~(b,e) with FFTs of the precipitate-containing regions presented in (c) and (f), respectively. For N=3 (c) the FFT is indexed to the \ce{Mg4Zn7} phase  (circles) and Mg matrix (squares). For N=20 (f) the FFT is indexed to the \ce{MgZn2} phase. A Hann filter was used to remove high frequency noise from the regions of interest in (b,e) before performing the FFT.
 \label{hrtem-23}}
\end{center}
\end{figure}

TEM observations on samples deformed to N=3 show narrow regions of stronger atomic contrast at the grain boundaries. These darker regions were typically 5--20\,nm in calliper diameter (see Fig.~\ref{hrtem-15}, in agreement with SAXS results) with widths of 5-10\,nm, no clearly defined facets, and extended along the grain boundary as a film. Figure~\ref{hrtem-23-a} shows a HRTEM image of the grain boundary region in this condition.  The two Mg grains present ($A$, left) and  ($C$, right)  are separated by a darker region  ($B$) at the grain boundary. A FFT of region $B$ is shown in Fig~\ref{hrtem-23-fft-Mg4Zn7} with reflections assigned to the  \ce{Mg4Zn7} phase. This phase generally forms in the shape of high aspect ratio rods in isothermally aged Mg-Zn alloys \cite{Gao2007,Singh2007}. \ce{Mg4Zn7} has been reported at grain boundaries in conventionally cast Mg-Zn alloys  \cite{GaoInter2007} where it is present in Mg--\ce{Mg4Zn7} lamellae. The beam direction, $\boldsymbol{B}$, is  parallel to the ${[}2\overline{4}2\overline{3}{]}$ zone axis in ($A$) and is close to the [010]  zone axis of \ce{Mg4Zn7} in ($B$). This orientation relationship is not among those previously reported  for this phase in Mg \cite{Singh2007} and lacks the usual alignment of the $[010]_\ce{Mg4Zn7}$ ($d=0.52$\,nm) parallel to $[002]_\ce{Mg}$ ($d=0.26$\,nm). The adjacent grain, (C), is not aligned along a rational, low-index direction. The insets show an enlargement of the matrix (left) and a simulated HRTEM image (right) for defocus=70\,nm, thickness=70\,nm.
The absence of  i) a favourable orientation relationship, ii) clear crystallographic facets and and iii) the usual rod-like morphology suggest that the precipitate nucleated at the grain boundary rather than intragranularly.  A full listing of the reflections used to index the phases in regions $(A, B)$ is set out in Tables~\ref{tab-fft-hrtem-23a} and \ref{tab-fft-hrtem-23b} in the supplementary information (SI).   

More extensive deformation results  in the grain-boundary precipitates coalescing into roughly equiaxed particles. A typical micrograph showing this condition is shown in Figure~\ref{N20_10}. The larger, dark regions are Mg grains in a strongly diffracting condition. Precipitates are concentrated at the grain boundaries and triple points, with two examples indicated by arrows. 
It was possible to obtain atomic resolution images of some of the precipitates, as shown in Figure~\ref{hrtem-n20}.
The Mg matrix (lower right) is close to a two beam condition with $\boldsymbol{g}=0002$. The precipitate (centre and left region)  has stronger atomic contrast.
Figure~\ref{hrtem-n20-fft} shows the FFT of the precipitate region which was assigned to the \ce{MgZn2} phase, with beam direction $[0001]$. A full listing of the reflections used to index the phase is set out in Table~\ref{tab-fft-_N20_6} in the SI.   

The TEM observations of N=3 samples suggest that the initial grain boundary precipitates are comprised of the \ce{Mg4Zn7} phase. Further solute diffusion of Zn to the grain boundaries continues during HPT deformation, resulting in the formation  of the \ce{MgZn2} phase. Although first principles calculations show \ce{MgZn2}  has a lower formation enthalpy for Zn $>$66at.\% \cite{Xie2013} the difference is minimal and microdomains of the \ce{Mg4Zn7} and \ce{MgZn2} phases can co-exist within individual precipitates  \cite{RosalieSingh2011,SinghRosalie2010,RosalieSomekawa2010} giving rise to a continuous spectrum of compositions. 
It is probable that such mixed-phase precipitates are present in the HPT material but extensive analysis would be required to confirm this.

\subsection{SAXS}

\begin{figure}
	\begin{center}
		\includegraphics[width=\textwidth]{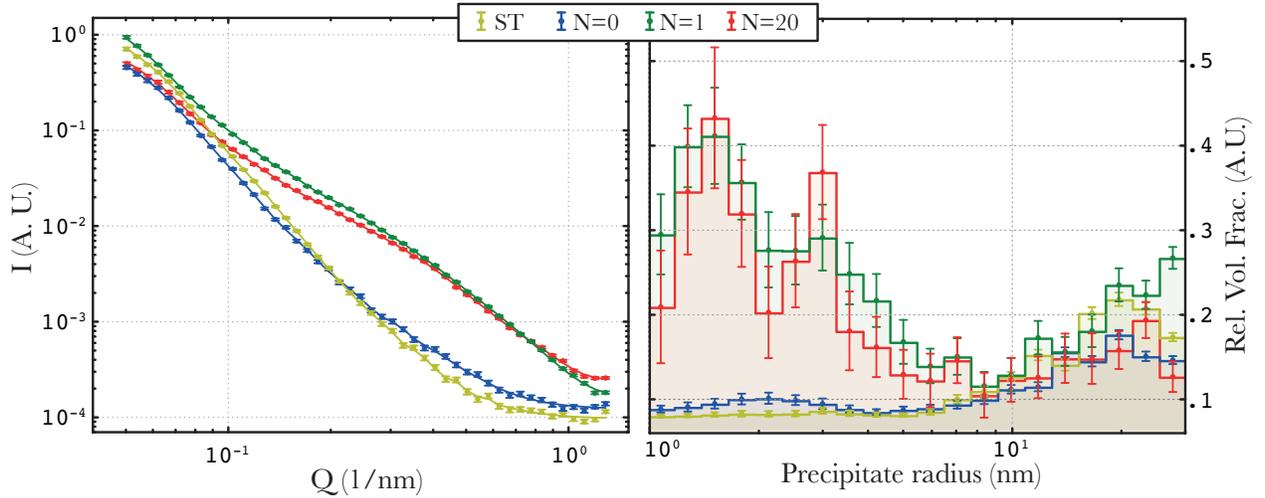}
		\caption{Left: Small-angle scattering data and fits (points with uncertainties, and solid lines, respectively). Right: Size distributions of scatterers determined from fitting of the scattering data. Error bars indicate $\pm$1\,standard deviation (SD). \label{fig-saxs}}
	\end{center}
\end{figure}

The scattering behaviour of the ST alloy shows little evidence for any structure in the measurable size range, and only shows the $I\propto Q^{-4}$ scattering from larger scattering structures (Fig.~\ref{fig-saxs}). After compression (N=0), there are slight but measurable changes in the scattering. Major changes occur upon HPT rotation, with strong, almost identical shoulders appearing in the scattering patterns for the N=1 and N=20 samples. 

The Monte Carlo method can retrieve a size distribution from the scattering patterns provided that a general scatterer shape is defined \cite{Pauw2013}. Given the diversity of grain boundary precipitate morphologies observed in TEM, a generally applicable spherical shape was chosen for the characterisation of the precipitate dimensions. Application of this method shows that the shoulders in the scattering patterns for both HPT samples can be interpreted as scattering originating from a broad distribution of objects, with radii ranging from about 2.5--20\,nm, and maximum volume-weighted radii of around 3--5\,nm. This size is consistent with the particle sizes observed via TEM. Due to technical limitations, the volume fractions shown in Fig.~\ref{fig-saxs} are not on absolute scale. They are, however, internally consistent and therefore allow for comparisons between samples  \footnote{Analysis of simulated data indicates that the determined volume fraction remains largely unchanged provided the aspect ratio of globular scatterers is less than 1:5.}. The volume fraction and size of detected precipitates in the N=1 and N=20 samples exhibit striking similarity. 

\section{Conclusions}
In conclusion, HRTEM and SAXS provided direct evidence of precipitation in an HPT-deformed Mg-3.4at.\%Zn alloy without post-deformation heat treatment as well as the characterisation of the precipitates. Both techniques show fine-scale precipitates after deformation with dimensions between 2.5--20\,nm, with the precipitate radii centred at around 3--5\,nm. HRTEM observations indicate that precipitation initially takes the form of a grain boundary film consisting  of the \ce{Mg4Zn7} phase.

SAXS shows little difference in the precipitate size and volume fraction in materials subjected to 20 rotations, where the alloy is in the the strain-saturated condition. However solute segregation to the grain boundaries continues during deformation, as evidenced by HRTEM showing a transition from a microstructure with a grain boundary \ce{Mg4Zn7} film to one with more equiaxed particles of the \ce{MgZn2} phase.

\section*{Author contributions}
The HPT samples were prepared by FM. JMR and FM performed the SEM and TEM investigations, with the analysis of the HRTEM micrographs being done by JMR. BRP performed the SAXS measurements, corrections and analysis. JMR and BRP wrote the majority of the manuscript, with contributions from FM and KT and with feedback and comments from all authors. HM and HK are responsible for the SAXS instrument used in the studies. KT supervised the investigations presented herein.

\section*{Acknowledgements}
The authors would like to thank Dr. H. Somekawa for supplying the extruded material used in the investigations. This work was supported in part by a Grant-in-Aid for Scientific Research on Innovative Area, "Bulk Nanostructured Metals", through MEXT, Japan (contract no. 22102004). 

\bibliography{References.bib}



\section*{Supplementary information}

\begin{table}[hbtp]
\begin{center}
\caption{The complete listing of the reflections used to index the Fast Fourier transforms of region (A) of the HRTEM image in Fig~\ref{hrtem-23-a}. Reflections are assigned to  \ce{Mg}, with beam direction, $\boldsymbol{B}=[2\overline{4}2\overline{3}]$. The table lists the measured ($d$)  and calculated  ($d^\prime$) interplanar spacings for a given reflection. The relative difference between these two values is shown by the value $|1-(d/d^\prime)|$. Interplanar angles are measured with respect to the $10\overline{1}0$ reflection (indicated by an asterisk). The difference between the measured and expected angles ($\Delta\theta$) is also provided.
\label{tab-fft-hrtem-23a}}
\begin{tabular}{c@{ }c@{ }c@{ }ccccrrr}
\toprule
	&		& 	&		&	\multicolumn{3}{c}{$d_{hkl}$}					&	\multicolumn{3}{c}{Angle}					\\ 
\cmidrule(r){5-7}\cmidrule(l){8-10} 
	&		&&		&	Meas.($d$)	&	Expect.($d^\prime$) &		&	Meas.	&	Expect.	&	$\Delta\theta$	\\
h	&	k	&	$\cdot$	&l	&	(\AA)	&	(\AA)		&	$|1-(d/d^\prime)|$	&		$(^\circ)$			&	$(^\circ)$				&	$(^\circ)$					\\ \midrule
1	&	0	&	$\overline{1}$	&	0*	&	0.279	&	0.277	&	0.01	&	0.0	&	0.0	&	0.0	\\
$\overline{1}$	&	0	&	1	&	0	&	0.271	&	0.277	&	0.02	&	179.7	&	180.0	&	0.3	\\
1	&	$\overline{1}$	&	0	&	2	&	0.188	&	0.190	&	0.01	&	70.3	&	70.0	&	0.3	\\
0	&	$\overline{1}$	&	1	&	2	&	0.188	&	0.190	&	0.01	&	110.9	&	110.0	&	0.9	\\
$\overline{1}$	&	1	&	0	&	$\overline{2}$	&	0.189	&	0.190	&	0.00	&	110.1	&	110.0	&	0.1	\\
0	&	1	&	$\overline{1}$	&	$\overline{2}$	&	0.189	&	0.190	&	0.00	&	70.4	&	70.0	&	0.4	\\
1	&	1	&	$\overline{2}$	&	$\overline{2}$	&	0.135	&	0.136	&	0.01	&	137.8	&	137.6	&	0.3	\\
$\overline{1}$	&	$\overline{1}$	&	2	&	2	&	0.136	&	0.136	&	0.00	&	42.8	&	42.4	&	0.4	\\
2	&	0	&	$\overline{2}$	&	0	&	0.140	&	0.139	&	0.01	&	0.0	&	0.0	&	0.0	\\
$\overline{2}$	&	0	&	2	&	0	&	0.138	&	0.139	&	0.01	&	180.0	&	180.0	&	0.0	\\
\bottomrule
\end{tabular}
\end{center}
\end{table}

\begin{table}
\begin{center}
\caption{A complete listing of the reflections used to index the Fast Fourier transforms of region (B) of the HRTEM image in Fig~\ref{hrtem-23-a}. Reflections are assigned to the  \ce{Mg4Zn7} phase, with beam direction, $\boldsymbol{B}=[010]$. The table lists the measured ($d$)  and calculated  ($d^\prime$) interplanar spacings for a given reflection. The relative difference between these two values is shown by the value $|1-(d/d^\prime)|$. Interplanar angles are measured with respect to the 206 reflection (indicated by an asterisk). The difference between the measured and expected angles ($\Delta\theta$) is also provided.\label{tab-fft-hrtem-23b}}
\begin{tabular}{c@{ }c@{ }c@{ }ccccrrr}
\toprule
	&		&		&	\multicolumn{3}{c}{$d_{hkl}$}					&	\multicolumn{3}{c}{Angle}					\\ 
\cmidrule(r){4-6}\cmidrule(l){7-9} 
	&		&		&	Meas.($d$)	&	Expect.($d^\prime$	)&		&	Meas.	&	Expect.	&	$\Delta\theta$	\\
h	&	k	&	l	&	(\AA)	&	(\AA)		&	$|1-(d/d^\prime)|$	&		$(^\circ)$			&	$(^\circ)$		 		&	$(^\circ)$					\\ \midrule

2	&	0	&	6*	&	0.222	&	0.228	&	0.00	&	0	&	0.0	&	0	\\
$\overline{4}$	&	0	&	6	&	0.222	&	0.244	&	0.06	&	33.2	&	32.2	&	1.0	\\
4	&	0	&	$\overline{6}$	&	0.220	&	0.244	&	0.07	&	148.1	&	147.8	&	0.3	\\
$\overline{2}$	&	0	&	$\overline{6}$	&	0.218	&	0.228	&	0.02	&	179.5	&	180.0	&	0.5	\\
12	&	0	&	$\overline{3}$	&	0.213	&	0.211	&	0.04	&	97.8	&	92.6	&	5.2	\\
$\overline{12}$	&	0	&	3	&	0.210	&	0.211	&	0.02	&	83.0	&	87.4	&	4.4	\\
8	&	0	&	5	&	0.178	&	0.194	&	0.05	&	23.4	&	26.6	&	3.2	\\
$\overline{8}$	&	0	&	$\overline{5}$	&	0.176	&	0.194	&	0.06	&	156.1	&	153.4	&	2.7	\\
$\overline{8}$	&	0	&	7	&	0.167	&	0.194	&	0.11	&	45.4	&	46.8	&	1.4	\\
8	&	0	&	$\overline{7}$	&	0.164	&	0.194	&	0.12	&	135.5	&	133.2	&	2.2	\\
12	&	0	&	0	&	0.192	&	0.211	&	0.06	&	63.7	&	67.4	&	3.6	\\
$\overline{12}$	&	0	&	0	&	0.191	&	0.211	&	0.07	&	116.6	&	112.6	&	3.9	\\
\bottomrule
\end{tabular}
\end{center}
\end{table}

\begin{table}
\begin{center}
\caption{A complete listing of the reflections used to index the Fast Fourier transforms of region (B) of the HRTEM image in Fig~\ref{hrtem-n20}. The reflections are assigned to the \ce{MgZn2} phase, with beam direction, $\boldsymbol{B}=[0001]$. The table lists the measured ($d$)  and calculated  ($d^\prime$) interplanar spacings for a given reflection. The relative difference between these two values is shown by the value $|1-(d/d^\prime)|$.     Interplanar angles are measured with respect to the $\overline{1}2\overline{1}0$ reflection  (indicated by an asterisk). The difference between the measured and expected angles ($\Delta\theta$) is also provided. \label{tab-fft-_N20_6}}
\begin{tabular}{c@{ }c@{ }c@{ }ccccrrr}
\toprule
	&		& 	&		&	\multicolumn{3}{c}{$d_{hkl}$}					&	\multicolumn{3}{c}{Angle}					\\ 
\cmidrule(r){5-7}\cmidrule(l){8-10} 
	&		& &		&	Meas.($d$)	&	Expect.($d^\prime$) &		&	Meas.	&	Expect.	&	$\Delta\theta$	\\
h	&	k	&	$\cdot$	&l	&	(\AA)	&	(\AA)		&	$|1-(d/d^\prime)|$	&		$(^\circ)$			&	$(^\circ)$				&	$(^\circ)$					\\ \midrule
$\overline{1}$	&	2	&	$\overline{1}$	&	0*	&	0.259	&	0.260	&	0.00	&	0.3	&	0.0	&	0.3	\\
1	&	1	&	$\overline{2}$	&	0	&	0.226	&	0.260	&	0.13	&	63.7	&	60.0	&	3.7	\\
2	&	$\overline{1}$	&	$\overline{1}$	&	0	&	0.227	&	0.260	&	0.13	&	116.5	&	120.0	&	3.5	\\
1	&	$\overline{2}$	&	1	&	0	&	0.252	&	0.260	&	0.03	&	177.1	&	180.0	&	2.9	\\
$\overline{1}$	&	$\overline{1}$	&	2	&	0	&	0.222	&	0.260	&	0.14	&	118.3	&	120.0	&	1.8	\\
$\overline{2}$	&	1	&	1	&	0	&	0.237	&	0.260	&	0.09	&	64.5	&	60.0	&	4.5	\\
0	&	3	&	$\overline{3}$	&	0	&	0.147	&	0.150	&	0.02	&	32.6	&	30.0	&	2.6	\\
3	&	0	&	$\overline{3}$	&	0	&	0.133	&	0.150	&	0.11	&	89.7	&	90.0	&	0.3	\\
3	&	$\overline{3}$	&	0	&	0	&	0.143	&	0.150	&	0.05	&	146.0	&	150.0	&	4.0	\\
0	&	$\overline{3}$	&	3	&	0	&	0.144	&	0.150	&	0.04	&	147.9	&	150.0	&	2.1	\\
$\overline{3}$	&	0	&	3	&	0	&	0.132	&	0.150	&	0.12	&	91.7	&	90.0	&	1.7	\\
$\overline{3}$	&	3	&	0	&	0	&	0.149	&	0.150	&	0.01	&	34.9	&	30.0	&	4.9	\\
$\overline{2}$	&	4	&	$\overline{2}$	&	0	&	0.130	&	0.130	&	0.00	&	5.5	&	0.0	&	5.5	\\
2	&	$\overline{4}$	&	2	&	0	&	0.130	&	0.130	&	0.00	&	178.5	&	180.0	&	1.5	\\
\bottomrule
\end{tabular}
\end{center}
\end{table}


\end{document}